\documentclass[final,5p,times,twocolumn,num]{elsarticle}

\usepackage{amssymb}

\usepackage[fleqn]{amsmath}
\usepackage{bm}
\usepackage{braket}
\usepackage{multirow}
\usepackage{booktabs}
\usepackage{siunitx}

\begin{document}

\begin{frontmatter}

\title{Intruder-driven mirror energy differences between $^{29}$Cl and $^{29}$Mg studied with antisymmetrized molecular dynamics}

\author[pnu,cens]{Dae Ik Kim}

\author[riken]{Masaaki Kimura}

\author[pnu]{Chang-Hwan Lee}

\author[cens]{Youngman Kim}
\ead{ykim@ibs.re.kr}

\affiliation[pnu]{%
    organization={Department of Physics, Pusan National University},
    city={Busan},
    postcode={46241},
    country={Korea}
}

\affiliation[cens]{%
    organization={Center for Exotic Nuclear Studies, Institute for Basic Science},
    city={Daejeon},
    postcode={34126},
    country={Korea}
}

\affiliation[riken]{%
    organization={RIKEN Nishina Center},
    city={Wako},
    postcode={351-0198},
    state={Saitama},
    country={Japan}
}

\begin{abstract}
To clarify the mirror energy differences (MEDs) of the proton-unbound nucleus $^{29}$Cl and their microscopic origins, 
we investigate the low-lying states of the $^{29}$Cl--$^{29}$Mg mirror pair using antisymmetrized molecular dynamics.
The calculation reasonably reproduces the normal and intruder states of $^{29}$Mg, 
while suggesting alternative spin-parity assignments for $^{29}$Cl.
The $1/2^+$ and $3/2^+$ states are predicted to form a nearly degenerate ground-state doublet with a small MED because of their similar intrinsic structures.
In contrast, the $3/2^-$ and $7/2^-$ intruder states exhibit large negative MEDs and are assigned to the observed resonances at approximately 500~keV and 1.1~MeV, respectively.
Their large MEDs originate from the reduced Coulomb energies associated with the stronger deformation and spatially extended proton distributions in the intruder configurations.
\end{abstract}


\end{frontmatter}

\section{\label{sec:introduction} Introduction}
Proton-rich nuclei near the drip line have attracted much attention in nuclear-structure studies. Owing to their weak binding and strong Coulomb effects, their shell structure, deformation, and valence-proton distributions can differ significantly from those of their mirror partners~\cite{Auerbach:1983,Ogawa:1999,Doornenbal:2007,Wimmer:2021yhi}. 
A well-known example is the different ground-state shapes reported for the $^{70}$Kr and $^{70}$Se pair~\cite{Wimmer:2021yhi}.

The mirror energy difference (MED) is a useful measure of charge-symmetry breaking in mirror excitation spectra~\cite{Lenzi2001PRL,Zuker2002PRL,Bentley2007PPNP}. 
It is defined as the difference in excitation energy between analog states in a mirror pair. The MED mainly reflects changes in the Coulomb energy associated with the size of the system and occupations of proton orbitals. 
It is well known that the spatial extension of a weakly bound $s$-wave proton can produce a large MED, which is called the Thomas–Ehrman shift~\cite{Thomas:1952tes,Ehrman:1951zz}. 
Large MEDs can also arise from changes in configuration. 
For example, in the $^{36}$Ca--$^{36}$S pair located near the $N=20$ island of inversion, the $2^+_1$ states exhibit an MED of $-246$ keV, mainly due to the reduced proton $s_{1/2}$ occupation in $^{36}$S~\cite{Doornenbal:2007,Burger2012,Valiente-Dobon:2018frm}. The intruder $0^+_2$ states exhibit an even larger magnitude of MED, $-516(130)$ keV~\cite{Valiente-Dobon:2018frm,Lalanne:2022pzu}, owing to the proton excitation across the $Z=20$ shell gap in $^{36}$Ca.

The $^{29}$Cl--$^{29}$Mg pair is particularly interesting in this context. 
$^{29}$Mg lies at the border of the $N=20$ island of inversion and has low-lying $pf$-shell intruder states~\cite{Terry:2008,Matta:2019yak,MacGregor:2021}.
Such intruder configurations are generally associated with enhanced deformation.
Its mirror nucleus $^{29}$Cl is proton unbound but accessible by in-flight decay spectroscopy~\cite{Mukha:2015fst,Xu:2017bet}. 
In Ref.~\cite{Mukha:2015fst}, the two lowest observed resonances in $^{29}$Cl were separated by approximately 500~keV and were tentatively assigned to the $1/2^+$ ground-state resonance and the $3/2^+$ first excited state. Note that $^{29}$Mg has the $3/2^+$ ground state and the $1/2^+$ first excited state separated by only 55~keV. 
The reversed ordering of the $1/2^+$ and $3/2^+$ states in $^{29}$Cl was proposed on the basis of a $^{28}\mathrm{S}+p$ potential-model estimate and was attributed to a large Thomas–Ehrman shift of the $s$-wave proton. 
However, since this interpretation relies on a single-particle estimate, it should be examined using a microscopic nuclear-structure model. Moreover, $^{29}$Mg has low-lying $3/2^-$ and $7/2^-$ intruder states, whose mirror partners may also exhibit sizable MEDs associated with their different intrinsic structures.

In this work, we investigate the low-lying normal and intruder states of the $^{29}$Cl--$^{29}$Mg mirror pair using antisymmetrized molecular dynamics combined with the generator coordinate method (AMD+GCM)~\cite{KANADAENYO2003497, Kimura:2003uf,Kimura:2016dku}.
We show that the MED of the $1/2^+$ state relative to the $3/2^+$ state is not large enough to account for the observed 500~keV spacing in $^{29}$Cl. Instead, the $3/2^-$ intruder state exhibits a large MED, and we propose that it corresponds to the observed 500~keV resonance.

\section{Framework of AMD+GCM}\label{sec:framework}
 We employ a microscopic Hamiltonian with the isospin-invariant Gogny D1S effective interaction~\cite{BERGER1991365} and the Coulomb interaction. 
 Therefore, charge-symmetry breaking in the present Hamiltonian originates only from the Coulomb interaction. 
 The isospin-breaking components of the nuclear interaction are neglected in the present calculation. 
 Although such terms can contribute to MEDs at the level of several tens of keV~\cite{Bentley2015PRC,Lenzi2020PRC}, 
 the large MEDs discussed below are dominated by the Coulomb energy associated with deformation and spatially extended proton configurations.

The intrinsic AMD wave function is represented by a Slater determinant of
single-particle Gaussian wave packets,
\begin{align}
\Phi_{\mathrm{int}}
&=\frac{1}{\sqrt{A!}}
\det\set{\varphi_i(\bm r_j)},\\
\varphi_i(\bm r)
&=\prod_{\sigma=x,y,z}
\left(\frac{2\nu_\sigma}{\pi}\right)^{1/4}
e^{-\nu_\sigma\left(r_\sigma-Z_{i\sigma}\right)^2}
\chi_i\xi_i.
\label{eq:amd}
\end{align}
For a given parity $\pi$, the parity-projected state is defined as
\begin{align}
\Phi_{\mathrm{int}}^\pi = P^\pi\Phi_{\mathrm{int}},
\quad
P^\pi=\frac{1+\pi P_r}{2},
\end{align}
where $P_r$ is the parity operator and $\pi=\pm1$.
The Gaussian centroids $\bm Z_i$, spin wave functions $\chi_i$, and Gaussian
widths $\nu_\sigma$ are optimized by energy variation of
$\Phi_{\mathrm{int}}^\pi$ under a constraint on the quadrupole deformation
parameter $\beta$, while $\xi_i$ denotes the proton or neutron isospin
function.

The parity-projected states obtained at different deformation parameters $\beta_i$ are projected onto eigenstates of the total angular momentum and superposed to construct the GCM ansatz as
\begin{align}
\Psi^{J\pi}_{M}=\sum_{Ki}f^{J\pi}_{Ki}P^J_{MK}
\Phi_{\mathrm{int}}^\pi(\beta_i),\label{eq:gcm}
\end{align}
where $P^J_{MK}$ is the angular-momentum projection operator.
The amplitudes $f^{J\pi}_{Ki}$ and the corresponding energies are obtained by solving the Hill--Wheeler equation~\cite{Hill1953hweq}.

To construct charge-symmetric GCM model spaces, the basis for $^{29}\mathrm{Mg}$ includes the intrinsic states optimized for $^{29}\mathrm{Mg}$ and those obtained by interchanging the proton and neutron labels of the intrinsic states optimized for $^{29}\mathrm{Cl}$. The basis for $^{29}\mathrm{Cl}$ is constructed analogously. Thus, the two GCM model spaces are mapped onto each other by proton-neutron interchange, eliminating artificial differences between the mirror spectra due to unequal model spaces.

Finally, to analyze the particle--core configurations, we calculate spectroscopic factors. Let $\Psi^{J\pi}_{M}(A)$ denote an AMD+GCM wave function of
$^{29}\mathrm{Mg}$ or $^{29}\mathrm{Cl}$, and let
$\Psi^{J_c\pi_c}_{M_c}(A-1)$ denote the lowest $J_c^{\pi_c}$
eigenstate of the $^{28}\mathrm{Mg}$ or $^{28}\mathrm{S}$ core.
Integrating over the internal coordinates of the core, the overlap function is expanded as
\begin{align}
\psi(\bm r)
&=\braket{\Psi^{J_c\pi_c}_{M_c}(A-1)|\Psi^{J\pi}_{M}(A)}\nonumber\\
&=\sum_{\ell jm}C^{JM}_{J_cM_cjm}
u^{J_c\pi_c}_{\ell j}(r)/r
\left[Y_{\ell}(\hat{\bm r})\chi_{1/2}\right]_{jm},
\label{eq:overlap}
\end{align}
where $\bm r$ is the valence nucleon coordinate relative to the core. The spectroscopic factor (S-factor) for the channel $J_c^{\pi_c}\otimes\ell_j$ is defined as
\begin{align}
S_{\ell j}^{J_c\pi_c} = \int_0^\infty \left| u^{J_c\pi_c}_{\ell j}(r) \right|^2 dr,
\label{eq:sfactor}
\end{align}
which quantifies the magnitude of each channel.

\section{Results and Discussion}

\begin{figure}[tb]
  \centering
  \includegraphics[width=\linewidth]{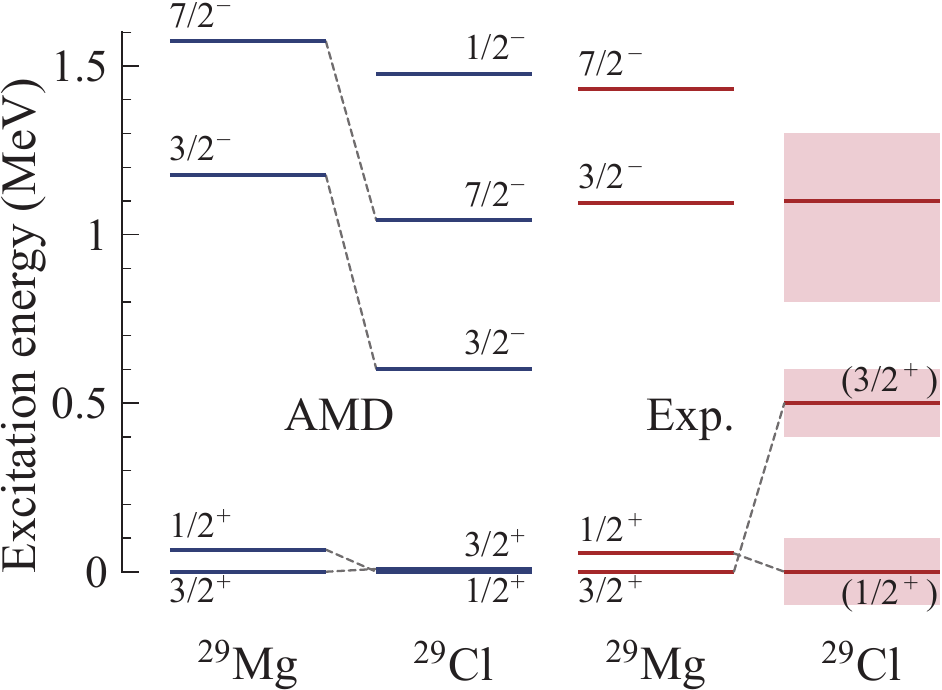}
  \caption{Excitation spectra of $^{29}$Mg and $^{29}$Cl obtained from the AMD+GCM calculation, compared with the experimental data.
  The experimental levels of $^{29}$Mg are taken from Ref.~\cite{Matta:2019yak}, and those of $^{29}$Cl from Ref.~\cite{Xu:2017bet}, where the spin-parity assignments in parentheses are tentative.}
  \label{fig:spectrum}
\end{figure}

We first discuss the low-lying states in $^{29}$Mg.
As shown in Fig.~\ref{fig:spectrum}, the calculation describes reasonably well both the positive-parity states with normal $sd$-shell configurations and the negative-parity states with $pf$-shell intruder configurations. 
It correctly reproduces the $3/2^+$ ground state and the very small excitation energy of the first $1/2^+$ state, as well as the excitation energies of the low-lying $3/2^-$ and $7/2^-$ intruder states.

Table~\ref{tab:sfac_mirror} clarifies the structure of these states. For the $1/2^+$ and $3/2^+$ states, the S-factors indicate their normal $sd$-shell configurations. Nevertheless, neither is a simple $s_{1/2}$ or $d_{3/2}$ single-particle state; the dominant component is coupled to the $2^+$ state of $^{28}$Mg rather than its ground state. The deformation of these states, $\beta \simeq 0.25$, induces strong particle--core coupling and gives rise to dominant core-excited components, a characteristic feature of nuclei in this mass region~\cite{Matta:2019yak,Urata:2011,Urata:2012,Macchiavelli:2017}.

The S-factors of the $3/2^-$ and $7/2^-$ states show that they have $pf$-shell configurations with sizable $p_{3/2}$ components. 
Note that the $p_{3/2}$ components indicate the neutron excitations across not only the $N=20$ but also the $N=28$ shell gap. In particular, the $3/2^-$ state is similar to the halo ground state of $^{31}$Ne~\cite{Nakamura:2009zzh,Hamamoto:2010,Minomo:2011bb,Urata:2011,Nakamura:2014,Takatsu:2022ubz} except for the neutron holes in the $sd$ shell. Because of their larger deformation with $\beta\simeq0.4$, the spectroscopic strength is fragmented among many core-excited channels including those omitted from Table~\ref{tab:sfac_mirror}. Both the proton and neutron root-mean-square (rms) radii are larger than those of the positive-parity states, which is also important in the discussion of their MEDs.

\begin{table*}[t]
\centering
\caption{Excitation energies, matter quadrupole deformations $\beta$, rms radii, and the S-factors of the low-lying states of the $^{29}$Cl--$^{29}$Mg mirror pair obtained with AMD+GCM. Each $J^{\pi}$ state is decomposed into the core nucleus $(J_c^{\pi_c})~\otimes$ valence-nucleon $(\ell_j)$ channels.
The experimental excitation energies of $^{29}$Cl are listed according to the spin-parity assignments proposed in the present work, rather than the tentative assignments of Refs.~\cite{Mukha:2015fst, Xu:2017bet}.
The MEDs quoted in the text and in Fig.~\ref{fig:med} are evaluated from the unrounded excitation energies.
}
\label{tab:sfac_mirror}
\sisetup{round-mode=places, round-precision=3, round-pad=true,
         table-format=1.2, detect-weight=true}
\setlength{\tabcolsep}{4.0pt}
\renewcommand{\arraystretch}{1.18}
\fontsize{9.0}{10.8}\selectfont
\begin{tabular*}{\textwidth}{@{\extracolsep{\fill}} l c S[round-precision=2,table-format=1.2] c *{3}{S[round-precision=2,table-format=1.2]} *{6}{S[round-precision=2,table-format=1.2]} @{}}
\toprule
 & {\multirow{2}{*}{$J^{\pi}$}} & \multicolumn{2}{c}{$E_{x}$ (MeV)}
   & {\multirow{2}{*}{$\beta$}} & \multicolumn{2}{c}{$r$ (fm)}
   & \multicolumn{6}{c}{S-factor} \\
\cmidrule(lr){3-4}\cmidrule(lr){6-7}\cmidrule(lr){8-13}
 & & {AMD} & {Exp.}
   & & {proton} & {neutron}
   & {$0^{+}\!\otimes\! s_{1/2}$} & {$2^{+}\!\otimes\! s_{1/2}$}
   & {$0^{+}\!\otimes\! d_{3/2}$} & {$2^{+}\!\otimes\! d_{3/2}$}
   & {$2^{+}\!\otimes\! d_{5/2}$} & {$4^{+}\!\otimes\! d_{5/2}$} \\
\midrule
$^{29}$Mg & $1/2^{+}$ & 0.066 & {0.055} & 0.248 & 3.072 & 3.200 & 0.42664 & {--} & {--} & 0.60223 & 0.05860 & {--} \\
$^{29}$Cl & $1/2^{+}$ & {0} & {0} & 0.255 & 3.212 & 3.083 & 0.47405 & {--} & {--} & 0.60139 & 0.06088 & {--} \\
\addlinespace
$^{29}$Mg & $3/2^{+}$ & {0} & {0} & 0.252 & 3.065 & 3.194 & {--} & 0.50080 & 0.32568 & 0.14043 & {--} & 0.09710 \\
$^{29}$Cl & $3/2^{+}$ & 0.008 & {0} & 0.256 & 3.200 & 3.071 & {--} & 0.58399 & 0.31821 & 0.13131 & {--} & 0.12515 \\
\midrule
 & {$J^{\pi}$} & {AMD} & {Exp.}
   & {$\beta$} & {proton} & {neutron}
   & {$0^{+}\!\otimes\! p_{3/2}$} & {$2^{+}\!\otimes\! p_{3/2}$}
   & {$4^{+}\!\otimes\! p_{3/2}$} & {$0^{+}\!\otimes\! f_{7/2}$}
   & {$2^{+}\!\otimes\! f_{7/2}$} & {$4^{+}\!\otimes\! f_{7/2}$} \\
\midrule
$^{29}$Mg & $3/2^{-}$ & 1.176 & {1.095} & 0.389 & 3.098 & 3.270 & 0.23450 & 0.13367 & {--} & {--} & 0.36838 & 0.05983 \\
$^{29}$Cl & $3/2^{-}$ & 0.601 & {0.500} & 0.396 & 3.282 & 3.109 & 0.23733 & 0.14311 & {--} & {--} & 0.37947 & 0.06657 \\
\addlinespace
$^{29}$Mg & $7/2^{-}$ & 1.574 & {1.431} & 0.378 & 3.090 & 3.260 & {--} & 0.22927 & 0.03704 & 0.28605 & 0.20542 & 0.05574 \\
$^{29}$Cl & $7/2^{-}$ & 1.043 & {1.100} & 0.385 & 3.269 & 3.098 & {--} & 0.25128 & 0.04234 & 0.28636 & 0.21175 & 0.05824 \\
\bottomrule
\end{tabular*}
\end{table*}

\begin{figure}[t]
  \centering
  \includegraphics[width=\linewidth]{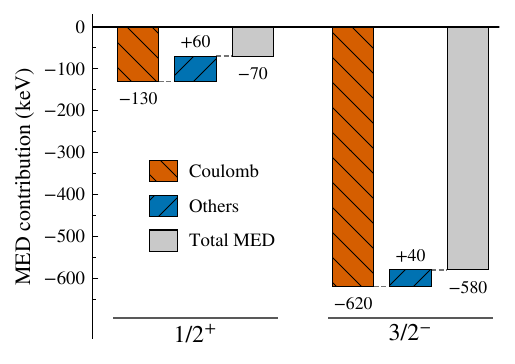}
  \caption{%
    Hamiltonian term-by-term decomposition of the mirror energy differences for the $1/2^{+}$ and $3/2^{-}$ states in ${}^{29}\mathrm{Cl}$ and ${}^{29}\mathrm{Mg}$.
    MED is evaluated with respect to the $3/2^{+}$ state for each nucleus, and decomposed into the Coulomb term and the remaining terms, kinetic energy + $NN$ interaction.
    }
  \label{fig:med}
\end{figure}

We now turn to $^{29}$Cl and discuss its MEDs.
The calculated spectrum shows a clear parity dependence: the positive-parity states have small MEDs, whereas the negative-parity intruder states have much larger negative MEDs. The $1/2^+$ state is slightly lowered below the $3/2^+$ state and becomes the ground-state resonance of $^{29}$Cl. However, its MED is much smaller than that suggested in Refs.~\cite{Mukha:2015fst,Xu:2017bet}, leaving the two positive-parity states nearly degenerate. In contrast, the $3/2^-$ and $7/2^-$ intruder states are lowered by approximately 500~keV.

Thus, we propose that the observed ground-state resonance contains an unresolved doublet of the nearly degenerate $1/2^+$ and $3/2^+$ states. The resonance at approximately 500~keV is assigned to the $3/2^-$ intruder state rather than the $3/2^+$ state, while the resonance at approximately 1.1~MeV may correspond to the $7/2^-$ intruder state.

Let us define the MED for quantitative discussion. Conventionally, it is defined as the difference between the excitation energies measured from the analog ground states. In the present mirror pair, however, the ground-state ordering is reversed. 
Therefore, throughout the following discussion, we evaluate the MED of each $J^\pi$ state using the $3/2^+$ state as a common reference,
\begin{align}
 {\rm MED}
 ={}& \left[E_{\rm Cl}(J^\pi)-E_{\rm Cl}(3/2^+)\right] \nonumber\\
 &\quad\quad\quad-\left[E_{\rm Mg}(J^\pi)-E_{\rm Mg}(3/2^+)\right].
\end{align}
Thus, the MED defined here differs from the conventional definition based on
the respective ground states and measures the energy shift of each state
relative to the common $3/2^+$ reference state;
a negative value means that the state is lowered in $^{29}$Cl relative to its mirror partner in $^{29}$Mg.
  
Figure~\ref{fig:med} shows the MEDs of the $1/2^+$ and $3/2^-$ states together with their decomposition. The MED of the $1/2^+$ state is only $-70$~keV, leaving it nearly degenerate with the $3/2^+$ state. In contrast, the MED of the $3/2^-$ state is $-580$~keV, almost an order of magnitude larger, and is almost entirely generated by the Coulomb contribution. The remaining contributions from the kinetic and nuclear-interaction terms are small and positive for both states. Since these terms are charge symmetric in the present calculation, their nonzero contributions reflect a slight Coulomb-induced rearrangement of the nuclear structure.

The different MEDs can be understood from the structural properties listed in Table~\ref{tab:sfac_mirror}. We first consider the positive-parity states. As in $^{29}$Mg, the $1/2^+$ and $3/2^+$ states of $^{29}$Cl exhibit typical particle-rotor structures, in which the $s$- and $d$-wave proton configurations are strongly coupled to core excitations. They can be regarded as members of the same $K^\pi=1/2^+$ rotational band and therefore have very similar intrinsic structures. Consistently, their proton rms radii are nearly identical, indicating similar Coulomb energies and hence a small MED between the two states.

A more detailed comparison of the S-factors reveals a structural rearrangement between the mirror nuclei. For both $1/2^+$ and $3/2^+$ states, the $s$-wave S-factors are slightly larger in $^{29}$Cl than in $^{29}$Mg. This enhancement reflects a Coulomb-induced rearrangement of the particle--core configuration.

The negative-parity states are also similar to their mirror partners. Their stronger deformation produces large core-excited components and distributes the S-factors over  the coupled $p$- and $f$-wave channels. In particular, the proton $p_{3/2}$ orbit is the mirror counterpart of the spatially extended neutron orbit responsible for the halo structure of $^{31}$Ne.
The combined effects of the stronger deformation and the spatially extended $pf$-shell proton components increase the proton radii of the negative-parity states by nearly 0.1~fm relative to the positive-parity states. The resulting reduction of the Coulomb energy is the origin of their large negative MEDs.

Thus, the pronounced parity dependence of the MEDs reflects the different collective structures of the normal and intruder states, rather than a simple single-particle picture.
For the positive-parity states, their similar particle-rotor structures and proton radii result in similar Coulomb energies, and hence a small MED.
For the negative-parity intruder states, the stronger deformation and more spatially extended proton distributions reduce the Coulomb energies and produce much larger negative MEDs.
The MEDs thus provide a sensitive probe of intruder configurations in the proton-rich mirror of the island of inversion.

\begin{figure}[t]
  \centering
  \includegraphics[width=\linewidth]{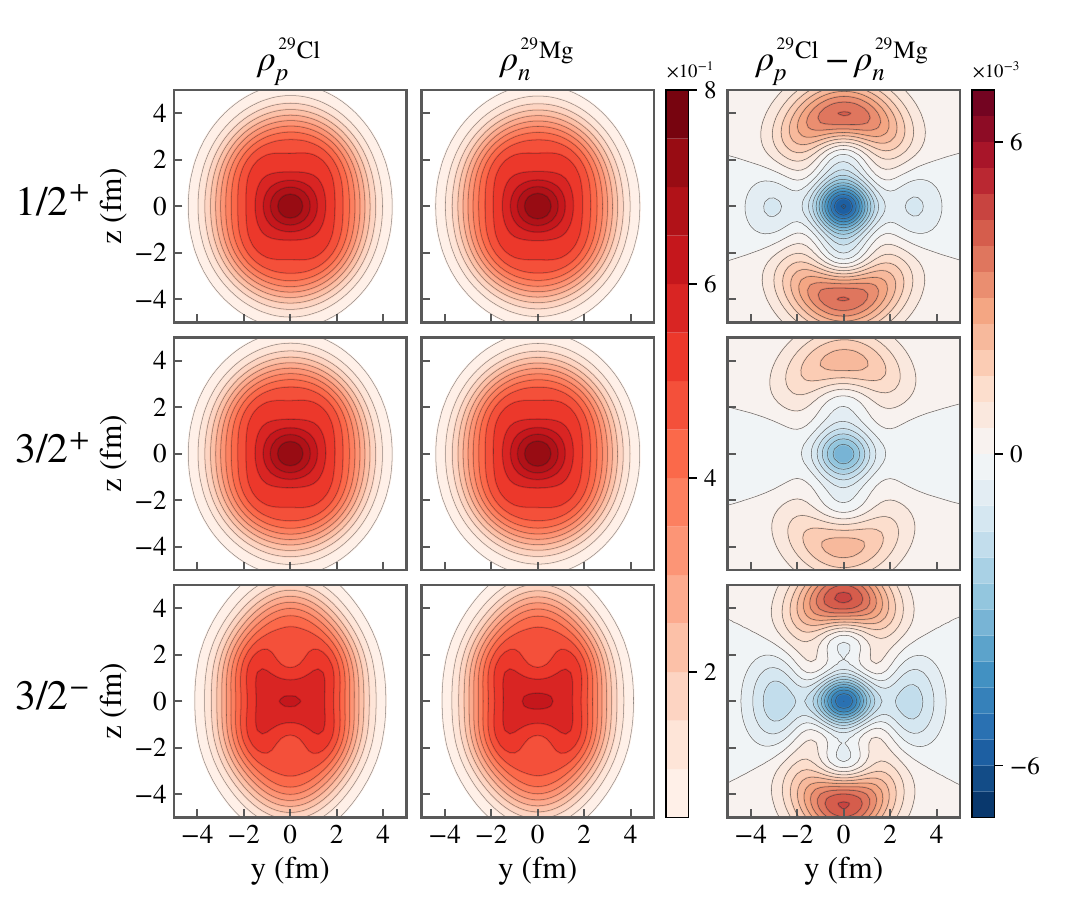}
  \caption{Overlap-weighted intrinsic density distributions of the $1/2^{+}$, $3/2^{+}$, and $3/2^{-}$ states (see text). The left and middle columns show the proton density of $^{29}$Cl and the neutron density of $^{29}$Mg in the $y$--$z$ plane, which would be identical under exact mirror symmetry. The right column shows their difference, $\rho_{p}(^{29}\mathrm{Cl})-\rho_{n}(^{29}\mathrm{Mg})$, on a different color scale, indicating the Coulomb-induced rearrangement. All densities are given in units of the saturation density $\rho_0=0.16$~fm$^{-3}$.}
  \label{fig:contour}
\end{figure}

Finally, we visualize the Coulomb-induced rearrangement between the mirror partners in Fig.~\ref{fig:contour}. 
To visualize the intrinsic structure of each GCM state, we define an overlap-weighted intrinsic density as
\begin{align}
 \rho_{\mathrm{int}}^{J\pi}(\bm{r}) = 
 \frac{\sum_i o_i^{J\pi} \rho_{i}(\bm{r})}{\sum_i o_i^{J\pi}},
 \label{eq:intrinsic_density}
\end{align}
where the intrinsic density of each basis state is
\begin{align}
 \rho_{i}(\bm{r})
 = \frac{\bra{\Phi_{\mathrm{int}}(\beta_i)}
 \hat{\rho}(\bm{r})
 \ket{\Phi_{\mathrm{int}}(\beta_i)}
 }{
 \braket{\Phi_{\mathrm{int}}(\beta_i)|
 \Phi_{\mathrm{int}}(\beta_i)}
 }.
\end{align}
For each intrinsic wave function, the intrinsic density is evaluated in the body-fixed frame defined by the principal axes of its quadrupole deformation.
The weight is determined from the overlap between the GCM state and each angular-momentum-projected basis state,
\begin{align}
 o_{iK}^{J\pi}
 = \frac{ \left|
 \braket{\Psi_M^{J\pi}|P^{J}_{MK}\Phi^{\pi}_{\mathrm{int}}(\beta_i)}
 \right|^2
 }{
 \braket{P^{J}_{MK}\Phi^{\pi}_{\mathrm{int}}(\beta_i)|P^{J}_{MK}\Phi^{\pi}_{\mathrm{int}}(\beta_i)}}.
\end{align}
Because the overlap depends on $K$, for each intrinsic basis state we select the $K$ component having the largest overlap:
\begin{align}
 o_i^{J\pi}
 &=
 o_{iK_i}^{J\pi},\quad
 K_i =
 \underset{K=-J,\ldots,J}{\operatorname{arg\,max}}\,
 o_{iK}^{J\pi}.
 \label{eq:intrinsic_density_weight}
\end{align}

The left and middle columns show the mirror-conjugate densities, $\rho_p$ of $^{29}$Cl and $\rho_n$ of $^{29}$Mg, respectively. Under exact mirror symmetry, the two densities would be identical.
Their small difference is shown in the right column. It remains at the percent level but shows a systematic outward shift of the proton density in $^{29}$Cl relative to the neutron density in $^{29}$Mg. This small rearrangement is also reflected in the slightly larger proton radii of $^{29}$Cl and the changes in the S-factors listed in Table~\ref{tab:sfac_mirror}. Its small magnitude indicates that it provides only a secondary correction to the MEDs. The main contrast between the positive- and negative-parity MEDs originates from their different configurations and spatial extensions, which are already present without this rearrangement.

\section{Summary and perspectives}
We have investigated the low-lying states of the $^{29}$Cl--$^{29}$Mg mirror pair using AMD+GCM to clarify the MEDs of the proton-unbound nucleus $^{29}$Cl and their microscopic origins.
We used an isospin-invariant nuclear interaction and constructed the charge-symmetric GCM model spaces so that differences between the mirror spectra arise solely from the Coulomb interaction.

The calculation reasonably reproduces the observed low-lying spectrum of $^{29}$Mg, including the nearly degenerate $3/2^+$ and $1/2^+$ normal states and the low-lying $3/2^-$ and $7/2^-$ intruder states.
We then examined the mirror counterparts of these states in $^{29}$Cl.
The calculations show a clear parity dependence of MEDs in $^{29}$Cl:
the positive-parity states have small MEDs, whereas the negative-parity intruder states exhibit large negative MEDs.
This result suggests new spin-parity assignments for $^{29}$Cl:
the ground-state resonance is interpreted as an unresolved doublet of the nearly degenerate $1/2^+$ and $3/2^+$ states, while the resonances at approximately 500~keV and 1.1~MeV are assigned to the $3/2^-$ and $7/2^-$ intruder states, respectively.

The small MED between the $1/2^+$ and $3/2^+$ states owes to their similar intrinsic structures, which can be interpreted as the members of the $K^\pi=1/2^+$ rotational band. Both states contain large core-excited components in which  $s$- and $d$-wave proton configurations are coupled to the core excitations, and their nearly equal proton radii lead to similar Coulomb energies. Consequently, the $1/2^+$ state is shifted by only $-70$~keV relative to the $3/2^+$ state. In contrast, the $3/2^-$ and $7/2^-$ intruder states have stronger deformation and more spatially extended proton distributions, which reduce their Coulomb energies and produce large negative MEDs. The Coulomb interaction also induces a slight structural rearrangement between the mirror partners, as reflected in the changes in the proton radii and spectroscopic factors, although this effect provides only a secondary contribution to the MEDs.

These results demonstrate that MEDs can probe intruder configurations in the proton-rich mirror of the island of inversion.
Extending the present framework to the neighboring $^{27}$Cl--$^{27}$Ne~\cite{Obertelli:2006wlh, Brown:2012zza, Grigorenko:2018cbw} and $^{31}$K--$^{31}$Mg~\cite{Neyens:2005zz, Kostyleva:2019dao} pairs will test how the MED evolves across the island and whether the island of inversion is mirror symmetric.

\section*{Acknowledgements}

D.I.K. and C.-H.L. were supported by the National Research Foundation of Korea (NRF) grant funded by the Korean government (No. RS-2023-NR076639). D.I.K. was supported by the Hyundai Motor Chung Mong-Koo Foundation.
M.K. acknowledges the support from Grant-in-Aid for Scientific Research (No.~JP26K00703).
Computational resources were partly provided by the National Supercomputing Center of Korea with supercomputing resources including technical support (KSC-2025-CRE-0562). This work was supported in part by the Institute for Basic Science, Korea (IBS-R031-D1) and by the National Research Foundation of Korea funded by Ministry of Science and ICT (RS-2024-00436392).

\end{document}